\documentclass[12pt]{spieman}  
\usepackage{amsmath,amsfonts,amssymb}
\usepackage{setspace} 
\usepackage{tocloft} 
\usepackage{graphicx} 
\usepackage{comment} 
\usepackage{fixltx2e} 
\usepackage{caption} 
\usepackage{subcaption} 
\usepackage[T1]{fontenc}
\usepackage[utf8]{inputenc}
\usepackage{xcolor}
\usepackage{textcomp}
\usepackage[ruled,vlined]{algorithm2e}
\usepackage[english]{babel}
\usepackage{csquotes}
\usepackage{hyperref}
\title{Automated real-time spectral characterization of phase-change tunable optical filters using a linear variable filter and infrared camera}

\author[a]{David Bombara}
\author[b]{Calum Williams}
\author[c]{Stephen Borg}
\author[c,d,*]{Hyun Jung Kim}
\affil[a]{University of Nevada, Reno, Department of Mechanical Engineering, 1664 N. Virginia Street, Reno, NV, USA, 89557}
\affil[b]{Department of Physics, Cavendish Laboratory, University of Cambridge, JJ Thomson Avenue, Cambridge, CB3 0HE, UK}
\affil[c]{NASA Langley Research Center, 1 Nasa Drive, Hampton, VA, USA, 23666}
\affil[d]{National Institute of Aerospace, 100 Exploration Way, Hampton, VA, USA, 23666}

\cftpagenumbersoff{figure}
\cftpagenumbersoff{table} 
\begin{document} 
\maketitle

\begin{abstract}
Actively tunable optical filters based on chalcogenide phase-change materials (PCMs) are an emerging technology with applications across chemical spectroscopy and thermal imaging. The refractive index of an embedded PCM thin film is modulated through an amorphous-to-crystalline phase transition induced through thermal stimulus. Performance metrics include transmittance, passband center wavelength (CWL), and bandwidth; ideally monitored during operation (in situ) or after a set number of tuning cycles to validate real-time operation. Measuring these aforementioned metrics in real-time is challenging. Fourier-transform infrared spectroscopy (FTIR) provides the gold-standard for performance characterization, yet is expensive and inflexible---incorporating the PCM tuning mechanism is not straightforward, hence in situ electro-optical measurements are challenging. In this work, we implement an open-source \textsc{Matlab}\textsuperscript{\tiny\textregistered}-controlled real-time performance characterization system consisting of an inexpensive linear variable filter (LVF) and mid-wave infrared camera, capable of switching the PCM-based filters while simultaneously recording in situ filter performance metrics and spectral filtering profile. These metrics are calculated through pixel intensity measurements and displayed on a custom-developed graphical user interface in real-time. The CWL is determined through spatial position of intensity maxima along the LVF's longitudinal axis. Furthermore, plans are detailed for a future experimental system that further reduces cost, is compact, and utilizes a near-infrared camera.
\end{abstract}


{\noindent \footnotesize\textbf{*Corresponding author: }Hyun Jung Kim, \linkable{hyunjung.kim@nasa.gov}}

\begin{spacing}{1}
\section{Introduction}
Optical bandpass filters are critical components utilized in a plethora of systems and applications, from fluorescence microscopy to remote sensing \cite{Bhargava2012, gat_imaging_2000, lichtman_fluorescence_2005}. These filters are designed to transmit only a certain band of wavelengths (passband) and block all others. Conventional optical bandpass filters are passive---offering discrete static passbands, arising from the interference of alternating index dielectric thin-films \cite{Macleod1986, lequime_tunable_2004}. There are a growing number of imaging applications requiring precise spectral filtering across a range of wavelengths (tunability), with motorized filter wheels typically utilized \cite{Hagen2013, Fei2014}. However, filter wheels are bulky, have limited spectral coverage, and offer slow switching speeds. 
In recent years, there have been significant research efforts toward cost-effective tunable optical bandpass filters that can operate across multiple wavebands and exhibit fast switching capability, based on nonstandard material platforms. Chalcogenide phase-change materials (PCMs), and their integration into micro-and nano-structured filter designs \cite{Wuttig2017, wang_optically_2016, williams_tunable_2020, julian_reversible_2020}, are emerging as an attractive candidate for such a tunable filter. These exotic materials exhibit a large refractive index shift across multiple wavebands through a reversible phase transition (amorphous-to-crystalline) initiated through external energy stimuli. When integrated into established or novel optical filter design schemes---such as Fabry-Perot interference-based or metasurface-based filters---PCMs can provide an optically active medium to tune the passband center wavelength (CWL) through refractive index switching \cite{williams_tunable_2020, julian_reversible_2020, Wuttig2017, wang_optically_2016}. 

\begin{figure*}[t]
    \centering
    \includegraphics[width = 0.95\linewidth]{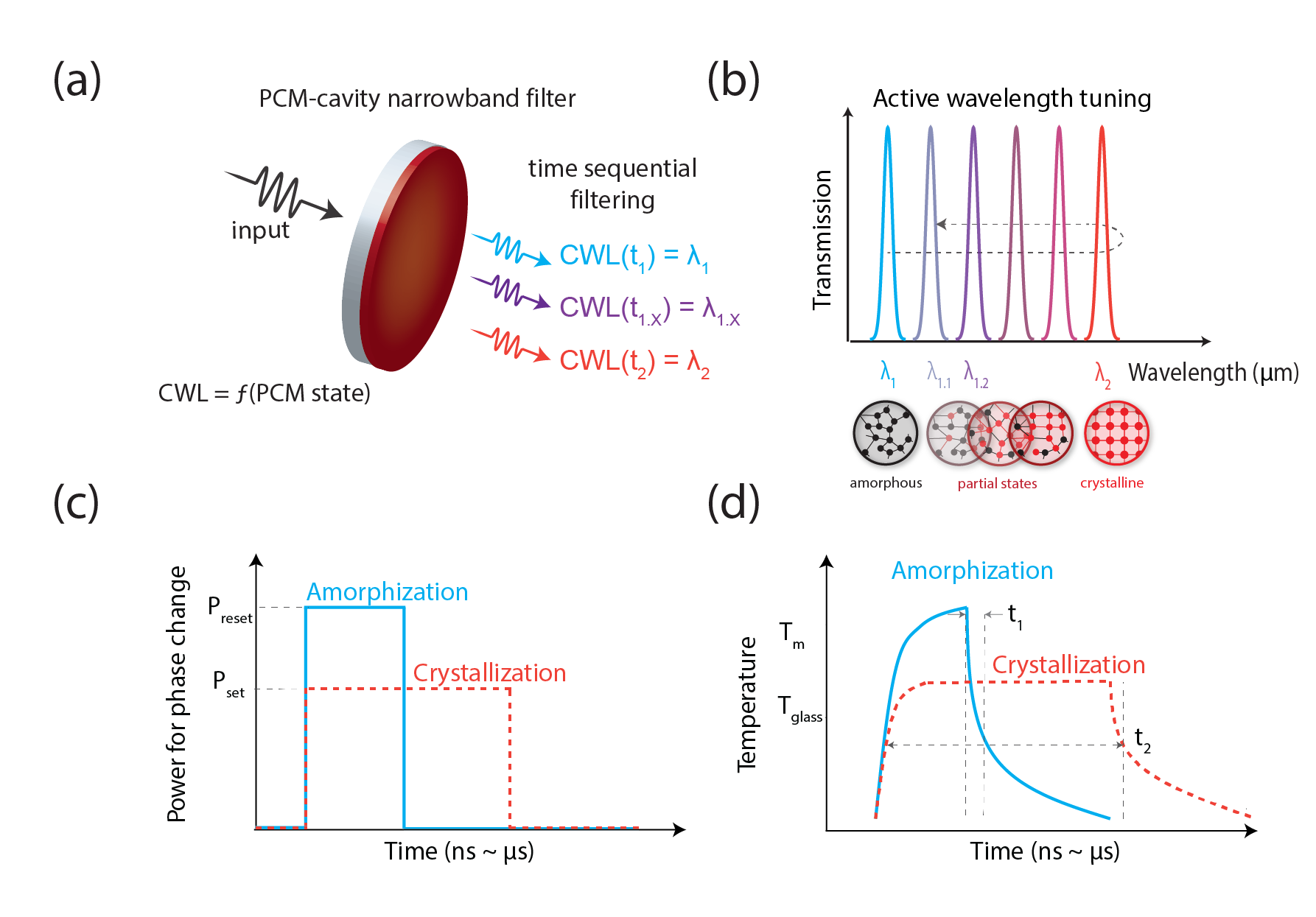}
    \caption{Active wavelength tuning of the phase-change material (PCM) cavity narrowband filter. (a) The PCM-cavity narrowband filter takes an optical input (illumination) and returns an optical output that is filtered at a corresponding center wavelength (CWL). The CWL is a function of the state of the PCM. (b) The CWL of the optical output is shortest when the PCM is in its amorphous phase, hence the lowest refractive index value. Partially crystalline states increase the refractive index, hence the CWL, until the PCM reaches the fully crystalline state (highest refractive index). (c) To ``reset'' the PCM to the amorphous phase, a pulse of high power but short duration must be applied (e.g., laser pulse). To set the material to its crystalline phase, a lower-power pulse of longer duration must be applied. (d) The material resets to its amorphous phase if it exceeds its melting temperature $T_m$, then is rapidly quenched below its glass transition temperature $T_{\mathrm{glass}}$ in time less than $t_1$. The crystalline transition requires that the PCM exceed $T_{\mathrm{glass}}$ for a duration of at least $t_2$.}
    \label{fig:pcm_overview}
\end{figure*}

PCMs undergo a phase transition through rapid localized melting and recrystallization \cite{Jafari2019}, which can be induced through optical or electrical stimuli. The former is typically used in rewritable optical storage media \cite{wuttig_phase-change_2007}. Compounds consisting of germanium, antimony, and tellurium (GeSbTe) are commonly used for optical devices, from optical storage media, integrated photonic elements, to tunable filters. To switch the material to its amorphous phase, an ultrafast (nanosecond), high-intensity pulse must be applied. In contrast, switching to the crystalline phase requires a longer, lower-intensity pulse. For amorphization, the material must be heated above its melting temperature, then rapidly quenched \cite{williams_tunable_2020}. For the material to crystallize, the material must be heated above its glass transition temperature for a duration long enough to reach crystallization. PCM-based tunable optical filters are typically switched using laser-based irradiation \cite{wuttig_phase-change_2007}. An overview of the PCM filter concept, laser input sequence, and corresponding temperatures of a PCM filter is shown in Fig. \ref{fig:pcm_overview}. Partial crystallization can also be realized, and intermediate phase switching has been previously shown to increase the number of accessible CWLs for a single multistate tunable filter \cite{julian_reversible_2020}.

PCM-based tunable filters have been shown to provide all-solid-state ultrafast (nanosecond) switching from the near-infrared (NIR) to mid-wave infrared (MWIR) \cite{julian_reversible_2020, williams_tunable_2020}. Despite the promising results, no studies, to the knowledge of the authors, have focused on device reliability metrics, which is crucial to understand  their behavior over many tuning cycles, device longevity, failure rates, and suitability for practical and real-world applications. For example, spectral properties that remain constant for each phase cycle would indicate that the filter functions reliably. However, when its properties begin to deviate from typical values, the filter may no longer be fit for its purpose. To determine these reliability metrics and characterize the tunability of the PCM-based filter, the spectral properties must be measured after each phase change. These include the passband CWL, transmittance intensity, and bandwidth (in terms of full width at half maximum, FWHM). Further, good homogeneity across the spatial extent of the filter is highly desirable, therefore the monitoring of this information is needed across the entire filter (i.e., 1D spatiospectral information as opposed to single point spectroscopic measurements). Fourier-transform infrared (FTIR) spectroscopy is arguably the most common method for obtaining this information and is the most accurate \cite{williams_tunable_2020, julian_reversible_2020}. For example, the Thermo Scientific\textsuperscript{\tiny\texttrademark} Nicolet\textsuperscript{\tiny\texttrademark} iS\textsuperscript{\tiny\texttrademark}10 FTIR spectrometer has a 10,000:1 signal-to-noise ratio and spectral resolution of 0.4\,cm$^{-1}$ \cite{thermo_advantages_2015}. However, it is expensive and bulky instrumentation that requires long data collection times to calculate the spectral properties from the interferogram. Critically, it is challenging to integrate the PCM-based filter switching systems within the FTIR apparatus and FTIR systems are generally limited to providing spectral properties from only a single point on the sample. 

In this study, we present an approach to real-time PCM-based tunable filter characterization that utilizes a linear variable filter (LVF) and IR camera to measure the filter's CWL. This method is simpler to construct for the research community and enables in situ measurement across the entire filter area (1D spatiospectral measurement). A graphical user interface (GUI) is developed in \textsc{Matlab}\textsuperscript{\tiny\textregistered} to control separate instruments and plot the tunable filter's passband CWL in real-time. These properties are measured based on acquired images from an IR camera. Furthermore, we detail progress on the PCM's reliability evaluation and a system for tuning and characterizing the PCM using commonly available, off-the-shelf hardware.
\section{Experimental Characterization} \label{section:characterization}
FTIR based analysis, albeit accurate, is expensive, time-consuming (especially if needed to perform spectral 2D measurements) and challenging to modify for our real-time performance characterization requirements. Instead, our alternate approach is to utilize a compact and inexpensive (relative to an FTIR spectrometer) LVF---a small rectangular optical filter that has its CWL varying linearly along the length of the filter \cite{sczupak_measurement_2018}. Therefore, in combination with the PCM-based filter, simply by imaging (recording) the intensity response through an LVF, it is possible to characterize the spectral performance. The accuracy of this method is high enough such that the phases of the PCM (amorphous and crystalline) are unambiguously distinguishable and the error margin of the CWL measurement is 0.085\,\textmu{}m. Error analysis is provided in Section \ref{section:exp-results}, ``Experimental Results.'' Our previous work realized a PCM-based filter with a CWL of approximately 2.9\,\textmu{}m and 3.3\,\textmu{}m in its amorphous and crystalline states, respectively \cite{julian_reversible_2020}. With this LVF-based system, the CWL shift of 0.4\,\textmu{}m is detectable. Further, this approach has the advantageous feature that it is simpler to modify in order to evaluate (test) other switching mechanisms or imaging approaches, such as trialing different electrode designs, different laser technologies, off-axis illumination, etc. 

\begin{figure*}
    \centering
    \includegraphics[width = 0.9\linewidth]{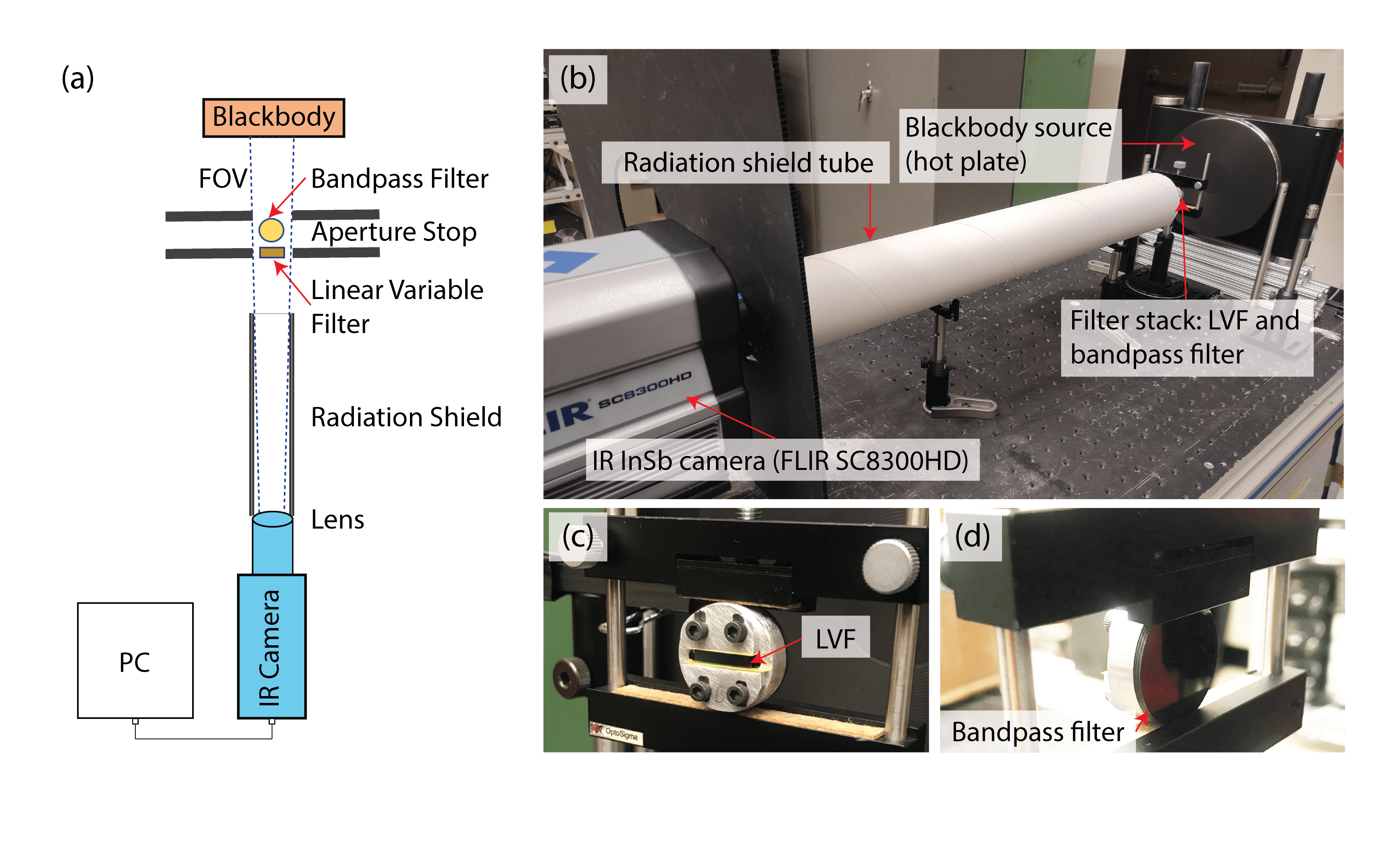}
    \caption{Experimental setup. (a) A schematic of the experimental setup. The infrared (IR) camera images the LVF with a fixed-wavelength bandpass filter behind it. They are both illuminated by the contact hot plate (blackbody). (b) A photograph of the experimental setup. The hot plate acts as a blackbody thermal radiation source. (c) The LVF is mounted in the IR camera's field of view. (d) The bandpass filter is between the blackbody source and LVF.}
    \label{fig:exp_setup}
\end{figure*}

\subsection{Experimental Setup}
To characterize a filter with tunable CWL, a system that validates the spectral performance of static bandpass filters is first needed. The experimental setup consists of (1) an IR camera (Indium antimonide detector, FLIR SC8000 Series), (2) a radiation shield tube, (3) a contact hot plate (blackbody radiation source), and (3) the filter stack (Fig. \ref{fig:exp_setup}). The IR camera is capable of imaging the MWIR waveband (2.5--5.0\,\textmu{}m) with a resolution of 1344 $\times$ 784 pixels. The camera utilized a lens with a focal length of 50\,mm. The distance between the lens and filter (target) was 122\,cm. The IR images are recorded using FLIR ResearchIR software. The blackbody radiation source was an iron plate with emissivity $\varepsilon \approx 1$ and a diameter of 17.8\,cm. A tube between the camera and filter stack shielded against unwanted background radiation.

A top-down schematic of the experimental setup is shown in Fig. \ref{fig:exp_setup}(a), whereas the photograph of the entire setup is shown in Fig. \ref{fig:exp_setup}(b). The LVF (Vortex Optical Coatings, Ltd.) and bandpass filter are shown in Fig. \ref{fig:exp_setup}(c) and \ref{fig:exp_setup}(d), respectively. The bandwidth of the LVF at 50\% peak transmittance is specified by the manufacturer to be 2\% of the peak wavelength with peak transmittance typically greater than 60\% across the CWL band \cite{noauthor_infra_nodate}. The LVF also has out-of-band blocking within its spectral range of an optical density of 3.0 \cite{noauthor_infra_nodate}. The LVF has nominal dimensions of 15\,mm\,$\times$\,3.5\,mm\,$\times$\,0.1\,mm. The CWL of the passband ranges from 2.5\,\textmu{}m to 5.0\,\textmu{}m, depending on the longitudinal distance along the filter.

The LVF is placed between the camera lens and as close to the bandpass filter as possible, minimizing any deviation due to converging/diverging beams. Similarly, the bandpass filter is placed between the LVF and the blackbody radiation source. The bandpass filters are manufactured by Andover Corporation and had CWL of 3.10\,\textmu{}m, 3.37\,\textmu{}m, 3.50\,\textmu{}m, 3.60\,\textmu{}m, 4.26\,\textmu{}m, 4.50\,\textmu{}m, and 4.70\,\textmu{}m. These filters each have a transmission of 75\% and an out-of-band blocking to an optical density of 3.0. The tolerance of the CWL of each filter is 40\,nm and each filter is 1.0\,mm\,$\pm$\,0.2\,mm thick. The bandwidths of the filters are between 120\,nm\,$\pm$\,30\,nm and 140\,nm\,$\pm$\,30\,nm.

\subsection{Experimental Results} \label{section:exp-results}
\begin{figure*}
    \centering
    \includegraphics[width = 0.99\linewidth]{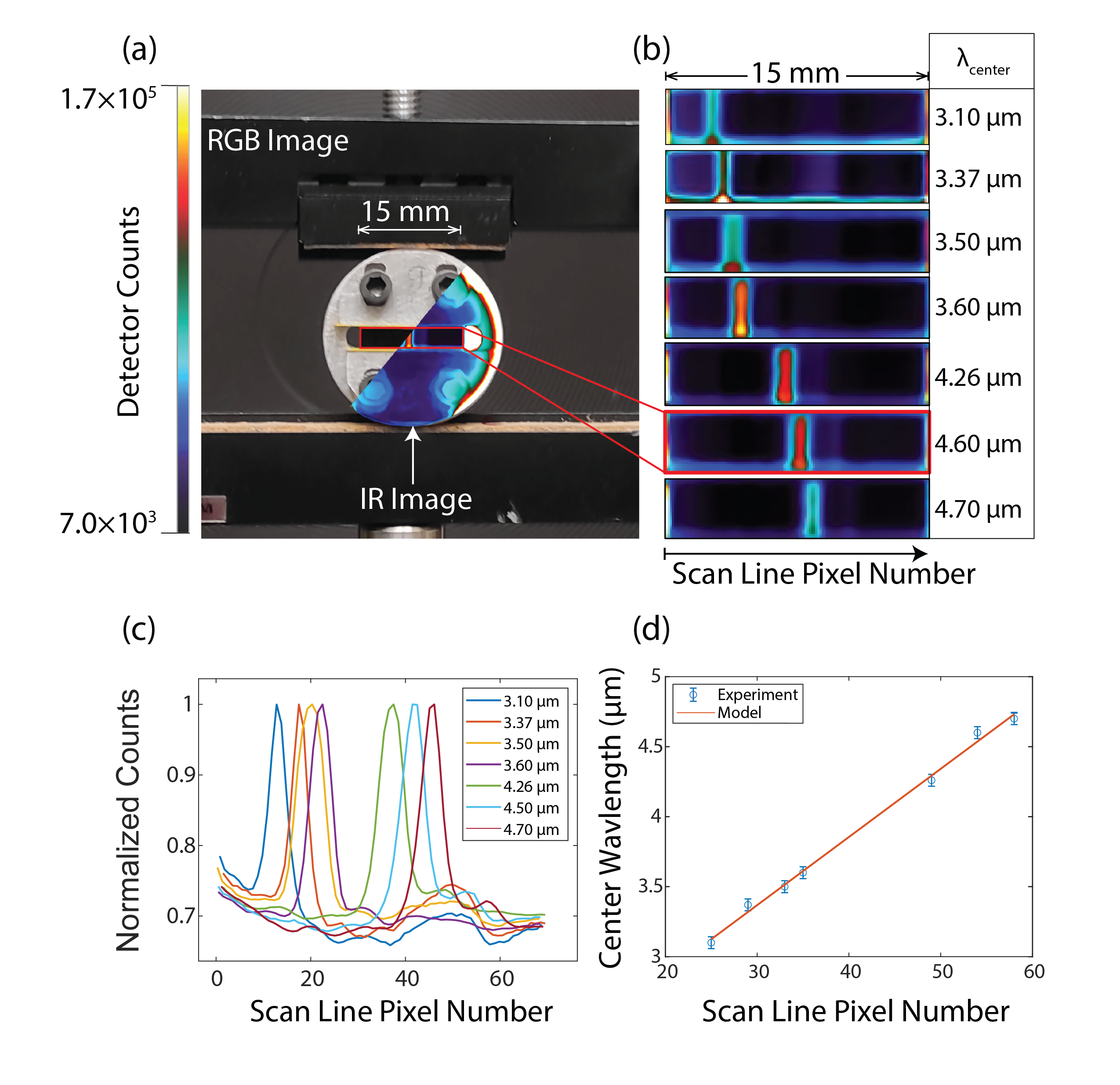}
    \caption{The results of the bandpass filter characterization using the LVF. (a) An IR image of the LVF overlayed on top of a red-green-blue (RGB) image of the filter stack that includes the LVF and bandpass filter. (b) The IR images of the LVF when bandpass filters of different CWL are placed behind it. (c) The normalized detector counts versus the scan line pixel number for each bandpass filter, with CWL shown in the legend. (d) The CWL versus scan line pixel number of the bandpass filter. The correlation is linear such that the scan line pixel number can be used to predict the CWL of the optical filter.}
    \label{fig:lvf_results}
\end{figure*}

Using the MWIR camera, the LVF was imaged with seven different bandpass filters behind it. Figure \ref{fig:lvf_results} shows the experimental results. The superimposed view of the IR intensity image and RGB image is shown in Fig. \ref{fig:lvf_results}(a) for illustration purposes. As shown in Fig. \ref{fig:lvf_results}, the CWL of the LVF varies along the length of the filter beginning at 2.5\,\textmu{}m and gradually increasing to 5.0\,\textmu{}m at the opposite end. However, due to variations in the manufacturing of our particular LVF, experimental data confirmed that it reached its 5\,\textmu{}m peak in just 10\,mm of filter length. The ``scan line pixel number'' in Fig. \ref{fig:lvf_results} refers to the numbering of the camera’s pixels along the horizontal axis of the LVF. The CWL here is defined to be the wavelength at which the transmittance is maximum. ``Detector counts'' from the IR camera are used as a proxy for transmittance; the location of peak detector counts is the location of the peak filter transmission. To find this location on the LVF, a single horizontal line of pixels is sampled in the center of the filter and parallel to the long axis. In Fig. \ref{fig:lvf_results}(b), the locations of peak transmission can be identified as regions of higher relative intensity, where the brightness of the pseudocolor image corresponds to the number of counts recorded during a given detector integration time. For each bandpass filter, the particular integration time was determined experimentally to obtain the high-contrast image and depended on the temperature of the blackbody radiation source.

In this experiment, we demonstrate that we can detect the CWL and accurately measure its location along the LVF. The plot of normalized detector counts versus scan line pixel number is shown in Fig. \ref{fig:lvf_results}(c) for each bandpass filter. The experimental results confirm that the LVF accurately indicates the CWL of the filter that is placed behind. The correlation between CWL and the scan line pixel number is shown in Fig. \ref{fig:lvf_results}(d). The pixel number of the scan line used as a proxy for the linear distance along the filter. 

For our experimental setup and detector/optics combination, the minimum horizontal spatial resolution for each pixel is approximately 0.341\,mm. The instantaneous field of view (IFOV) of our imaging system represents the spot size on the LVF for an individual pixel and also corresponds to our estimated spectral measurement uncertainty. Using a 50-mm lens in combination with a 1344$\times$784, 14-\textmu{}m-pixel-pitch sensor, we were able to sample the 15-mm LVF with approximately 104 pixels along the filter's horizontal (longitudinal) axis. As stated previously, the LVF did not use the full 15-mm length to cover the 2.5-\textmu{}m--5.0-\textmu{}m wavelength span. We measured the 2.5-\textmu{}m CWL shift on the LVF in approximately 10\,mm rather than 15\,mm. This slightly reduced the spectral resolution of the system. With our 10-mm (effective) LVF, we were seeing a 0.25-\textmu{}m/mm shift spatially along the filter rather than an expected 0.17-\textmu{}m/mm shift.

When combined with the IFOV of our imaging system (0.341\,mm), the spectral measurement uncertainty of our experiment can be calculated to approximately 0.085\,\textmu{}m. A 0.341\,mm spot size on the filter corresponds to a 0.085\,\textmu{}m change in wavelength across a pixel. In Fig. \ref{fig:lvf_results}(d), the CWL of the bandpass filter is the independent variable whereas the scan line pixel number is the dependent variable. If the CWL was unknown and the scan line pixel number was used to predict the CWL, the prediction error may be $\pm$0.085\,\textmu{}m. This is because an increase of 1\,mm along the LVF’s length corresponds to a 0.250\,\textmu{}m increase in the passband CWL. Similarly, an increase of 0.341\,mm (the horizontal IFOV) corresponds to an increase in the passband CWL of 0.250\,\textmu{}m/mm $\times$ 0.341\,mm = 0.085\,\textmu{}m. The IFOV is a function of the target distance and camera resolution. The measurement uncertainty may be minimized in future implementations through a change in imaging optics: for example a camera lens with a longer focal length, an image sensor with greater spatial resolution, and a longer (> 15-mm) LVF. All of these elements would improve our system accuracy by increasing the sampling of the LVF.

\begin{figure*}
    \centering
    \includegraphics[width = 0.99\linewidth] {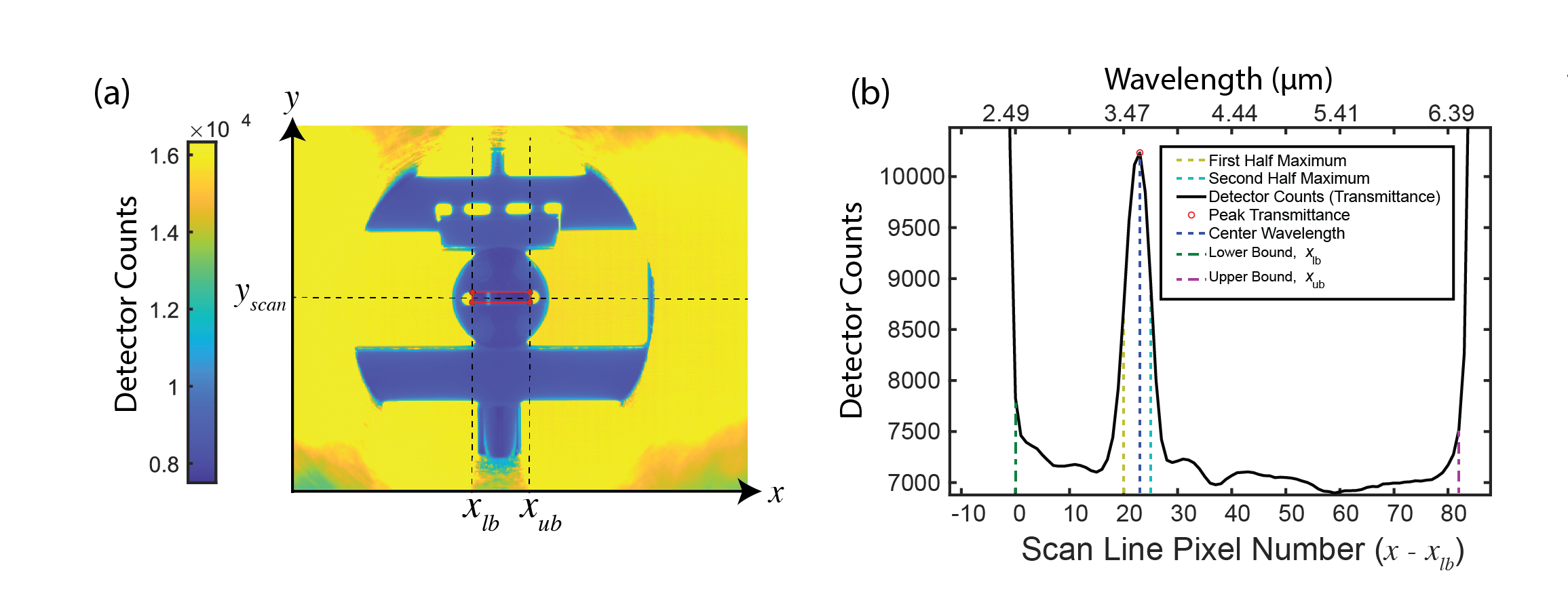}
    \caption{The method for determining the CWL of the filter from raw pixel data. (a) The IR image used to obtain CWL. The scan line is located at $y = y_{{scan}}$. The upper and lower bounds of the horizontal pixel number are $x_{lb}$ and $x_{ub}$, respectively. (b) The spectral results of the 3.60-\textmu{}m bandpass filter. Spectral properties shown include the locations of the CWL, peak transmittance, intensity half-maximums, and bounds of the filter.}
    \label{fig:lvf_explain}
\end{figure*}

\begin{figure}
    \centering
    \includegraphics[width = 0.8 \linewidth]{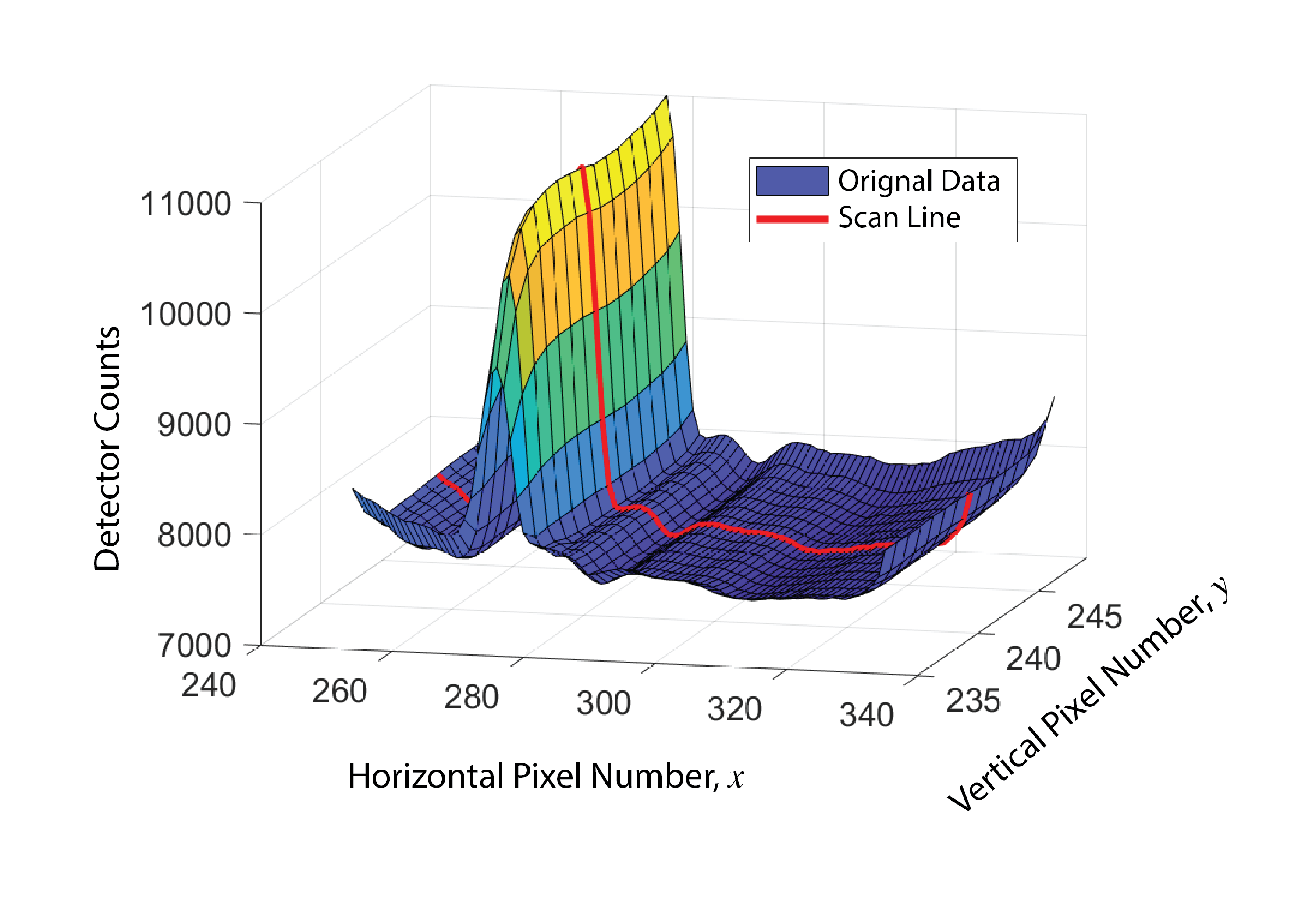}
    \caption{A three-dimensional plot showing the variation of the detector counts with the horizontal and vertical pixel numbers that correspond to the area of the IR image that the LVF occupies. As long as the scan line passes through the entire length of the LVF, the location of peak transmittance does not depend on the exact location of the scan line.}
    \label{fig:lvf_3d}
\end{figure}

The location of the scan line, $y_{scan}$, is shown in Fig. \ref{fig:lvf_explain}. To find the location of the CWL, first, a region of interest (ROI) is drawn over the filter image. $y_{scan}$ was chosen to be in the center of the LVF. It was initially assumed that the horizontal position of the peak transmittance may depend on the chosen location of $y_{scan}$, due to fabrication tolerances. However, Fig. \ref{fig:lvf_3d} shows otherwise, which shows the variation of the detector counts in the 2D pixel grid. Figure \ref{fig:lvf_3d} shows that the location of peak transmission along $x$ does not change depending on the chosen scan line location, $y_{scan}$. Another method to determine the location of peak transmittance would be to scan every pixel in the 2D ROI for every CWL measurement. Compared to scanning a single line, this method is less computationally efficient, making it less suited for the real-time characterization application. The number of detector counts is evaluated within two bounds, where $x_{lb}$ and $x_{ub}$ are the lower and upper bounds, respectively. In Fig. \ref{fig:lvf_explain}(a), $x$ and $y$ refer to the numbering of the image’s horizontal and vertical pixels, respectively. The detector counts must be evaluated within bounds because the detector counts increase dramatically outside the range of pixel numbers that do not cover the LVF. This increase can be seen in Fig. \ref{fig:lvf_explain}(b), where beyond $x_{lb}$ and $x_{ub}$ the colors that correspond to the detector counts appear brighter. Figure \ref{fig:lvf_explain}(b) shows results for the 3.60-\textmu{}m filter with labeled spectral properties. The locations of the CWL, FWHM, and peak transmittance are shown. The CWL is a function of the scan line pixel number; the two variables vary according to the equation,

\begin{equation}
    CWL = a(x - x_{lb}) + \lambda_0,
    \label{eq:linear-correlation}
\end{equation}
\noindent where $a$ is the proportionality constant, $\lambda_0$ is the passband CWL when $x = x_{lb}$, and $x$ is the numbering of the pixels horizontally. Least-squares linear regression identified $a = 0.04787\,\mathrm{\mu}\mathrm{m}$ and $\lambda_0 = 2.4917\,\mathrm{\mu}\mathrm{m}$. 

\section{Automated Performance Characterization}
\subsection{Overview}
\begin{figure*}[t]
    \centering
    \includegraphics[width = 0.9\linewidth]{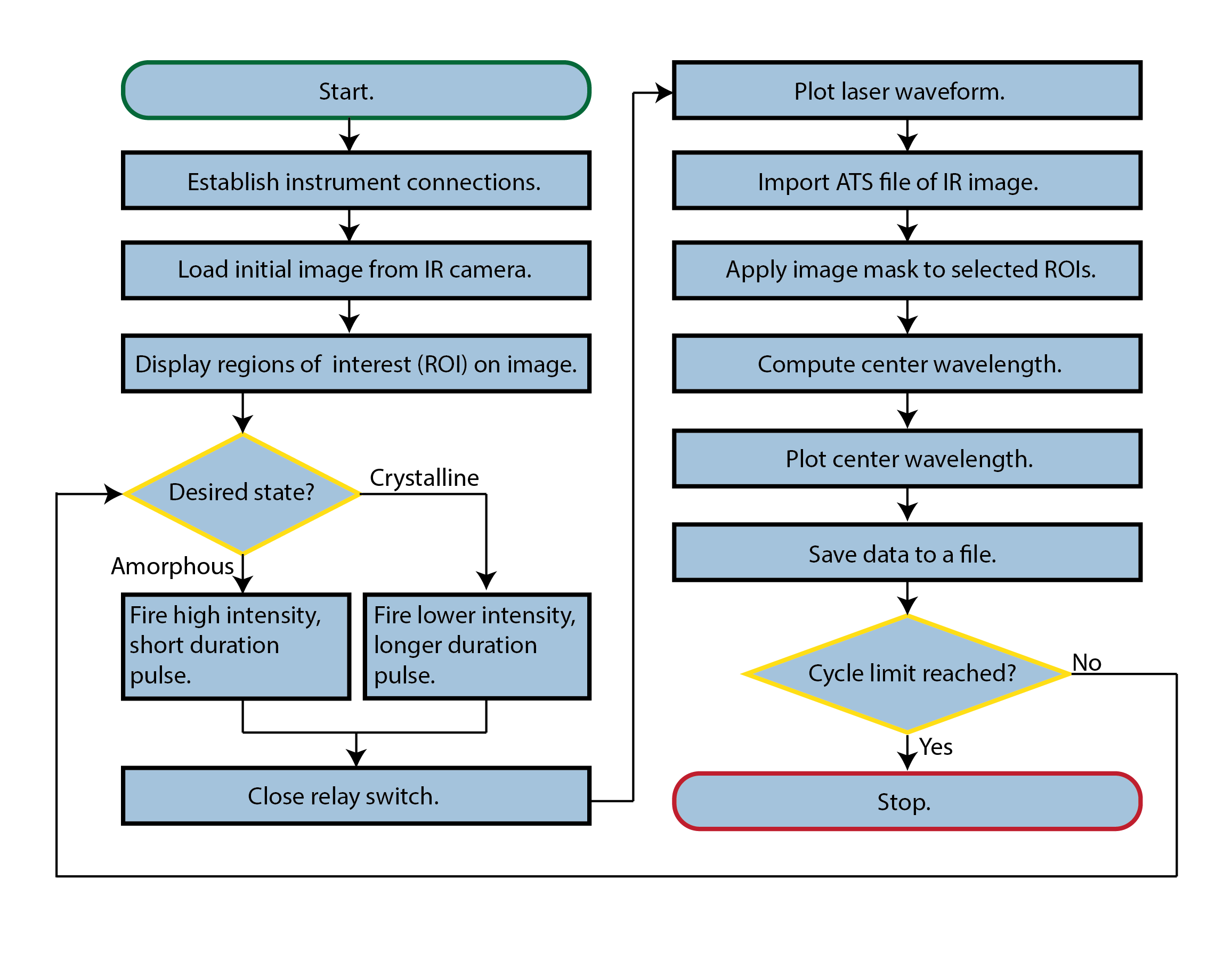}
    \setlength{\abovecaptionskip}{20pt plus 3pt minus 2pt} 
    \caption{Operation flowchart: The general commands and processes executed by the \textsc{Matlab}\textsuperscript{\tiny\textregistered} graphical user interface (GUI). The program runs in a loop until the cycle limit is reached but can be terminated by the user at any time.}
    \label{fig:code_flowchart}
\end{figure*}

The PCM tunable filters are based on thin-film Ge\textsubscript{2}Sb\textsubscript{2}Te\textsubscript{5} and fabricated in a laboratory at NASA Langley Research Center; the fabrication process is described in detail in other studies \cite{julian_reversible_2020, williams_tunable_2020, kim_active_2020}. The evaluation of tunable PCM-based filters may be automated in two aspects: phase switching and performance characterization. A \textsc{Matlab}\textsuperscript{\tiny\textregistered}-based application is developed to control the hardware systems and analyze incoming data. Figure \ref{fig:code_flowchart} details the main loop that runs in order to switch and characterize the PCM-based filter. The source code for the \textsc{Matlab}\textsuperscript{\tiny\textregistered} application is available at \url{https://dbombara.github.io/automated-pcm-characterization}, as well as a description of the experiment and setup.

Once the user begins the experiment, the filter automatically switches between the amorphous and crystalline states while its CWL is automatically calculated and saved for each cycle. The sequence of functions that automates the system is shown in Fig. \ref{fig:code_flowchart}. The GUI allows the user to conveniently control and monitor the experiment's progress. At the start of the experiment, \textsc{Matlab}\textsuperscript{\tiny\textregistered} establishes connections to each instrument. The initial IR image is then loaded, with the ROIs displayed on the screen. At this point, the user may redraw the ROIs, based on the exact position of the LVF in the view of the camera. The selected ROIs are displayed on the screen and remain in view while the experiment runs. Before firing a laser pulse, first, the desired state is evaluated. Typically, the desired state alternates on each cycle between amorphous and crystalline. The desired state of the PCM determines the intensity and duration of the fired laser pulse. A closed relay switch instructs the IR camera to save its image. The laser waveform is then plotted for that cycle. Then, the file of the IR image is imported and displayed in the GUI. An image mask is applied that allows the CWL to be calculated and plotted within two bounds on the horizontal axis. All data from that cycle are then saved to a file. The experiment runs until the cycle limit is reached, which can be preset by the user or determined by the GUI based on the data from each cycle. More details on these processes may be found below.

\subsection{Phase Switching}
The switching of the PCM between amorphous and crystalline states is triggered by the heat of an incoming laser pulse. A Quantel Evergreen HP pulsed laser is chosen in this study to deliver up to 340\,mJ of energy at 532\,nm center wavelength. The pulses are delivered at a rate of up to 20\,Hz. The wavelength and energy level were chosen based on our previous work on GeSbTe phase-change metasurface spectral filters \cite{julian_reversible_2020}. The pulsed laser output is controlled with a function generator and is viewed with an oscilloscope; this conveniently enables the user to monitor the waveform. The laser consists of two components: the power supply and the ``head'', which emits the pulsed laser beam. In this study, the oscilloscope (Tektronix MDO4024C) and function generator (Tektronix MDO4AFG) functionalities are combined into a single instrument (MDO stands for mixed-domain oscilloscope). This choice of oscilloscope and function generator was due to the high-level programming commands that are available in \textsc{Matlab}\textsuperscript{\tiny\textregistered}.

The laser, function generator, and oscilloscope are controlled within the \textsc{Matlab}\textsuperscript{\tiny\textregistered} user interface. From \textsc{Matlab}\textsuperscript{\tiny\textregistered}'s Instrument Control Toolbox\textsuperscript{\tiny\texttrademark}, Quick Control Oscilloscope commands are employed to issue high-level commands to the instruments. This aspect of the GUI builds upon the Oscilloscope App from MathWorks\textsuperscript{\tiny\textregistered} and incorporates similar functionality \cite{ursache_oscilloscope_2019}. Serial commands can also be issued directly to the laser to adjust the settings of the instruments, such as the power level \cite{rollefson_quantel_2018}.

\subsection{Performance Characterization}
The performance characterization is conducted in a way similar to the characterization in Section \ref{section:characterization}. To automate the image acquisition process, a USB-controllable relay (8-Channel 5-Amp ProXR Lite, Relay Pros) triggers the camera to record the IR images. In this study, the switch is ``dry''; the closure of the relay switch draws no electrical current. The FLIR ResearchIR software saves the image in a proprietary file format known as ATS. Then, the FLIR File Reader Software Development Kit (SDK) is used to import the ATS file into \textsc{Matlab}\textsuperscript{\tiny\textregistered}. Functions from the SDK then extract the image's metadata and raw pixel information so that the image can be displayed in the GUI. Using the imported image, the CWL of the tunable filter is calculated in the manner explained in Section \ref{section:characterization}, using Eq. \ref{eq:linear-correlation}.

\subsection{GUI Design}
\begin{figure*}[t]
    \centering
    \includegraphics[width = \linewidth]{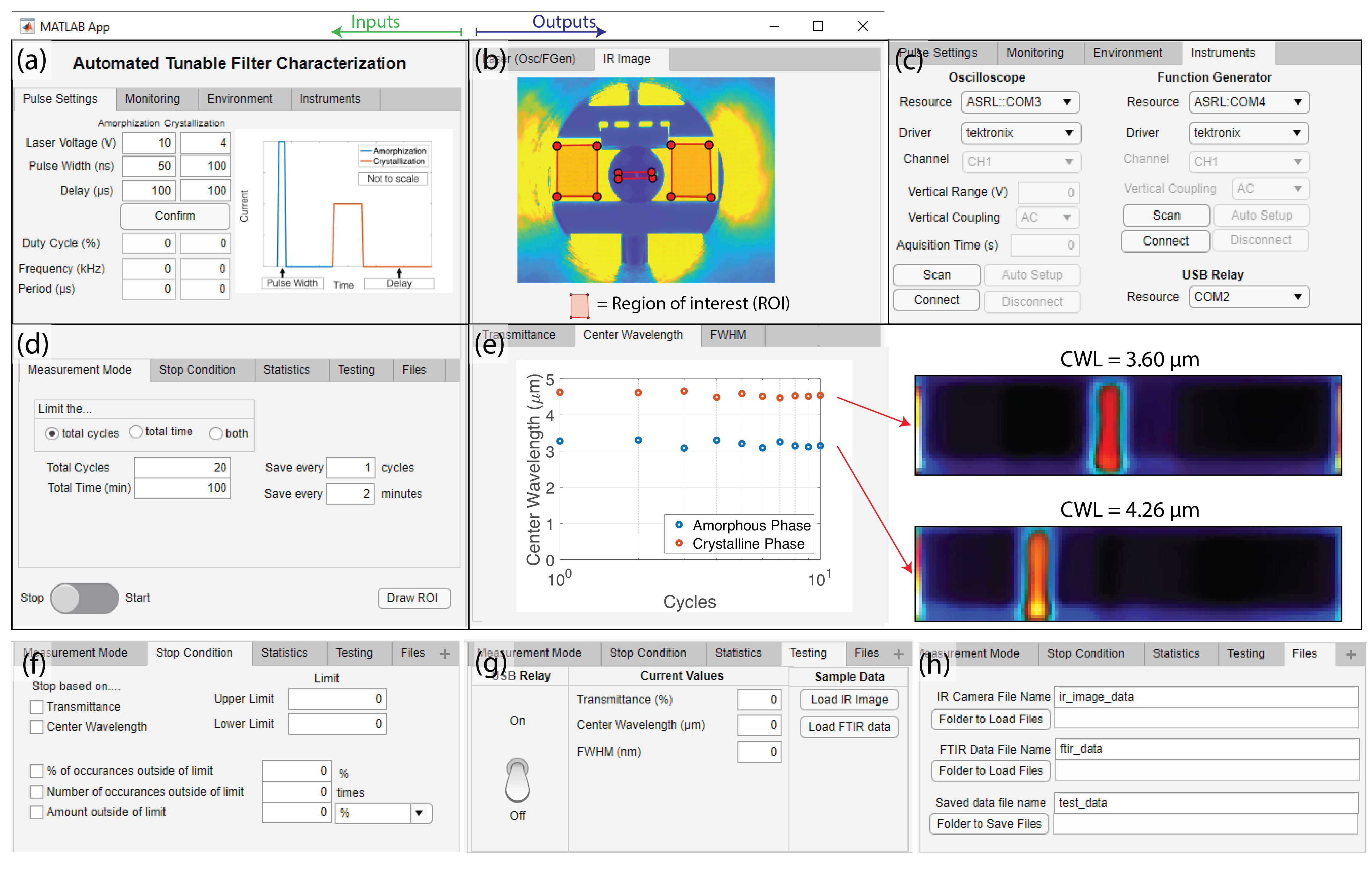}
    \setlength{\abovecaptionskip}{10pt plus 3pt minus 2pt}
    \caption{Real-time filter characterization: A screenshot of the custom \textsc{Matlab}\textsuperscript{\tiny\textregistered} GUI. The source code and packaged application are available at \url{https://dbombara.github.io/automated-pcm-characterization}. The window contains four quadrants (a)-(e), which each contain tabs. The left half of the GUI is generally for inputs by the user, whereas the right half displays the output from the instruments and computed spectral properties. (a) The top-left quadrant of the GUI contains tabs for hardware-based settings and conditions. (b) The top-right quadrant of the GUI displays ``raw'' data from the IR camera and laser. (c) The Instruments tab, which is a part of the top-left quadrant, enables the user to connect to the different instruments. (d) The bottom-left quadrant is for software-based settings. (e) The bottom-right quadrant displays the spectral properties of the tunable filter. With the current LVF-based approach, only the CWL may be plotted. In future work, other properties, such as transmittance and full width at half maximum (FWHM) will be calculated. The (f) Stop Condition tab, (g) Testing tab, and (h) Files tab allow the user to have more control over the experiment.}
    \label{fig:gui_overview}
\end{figure*}

The GUI is designed with the user's inputs on the left quadrants of the window and the system's outputs on the right quadrants. A screenshot of the GUI is shown in Fig. \ref{fig:gui_overview}. The various aspects of the GUI are organized with tabs. The names of the tabs include ``Instruments'', ``Measurement Mode'', ``Stop Condition'', ``Pulse Settings'', and ``Testing''. There are additional groups of tabs for monitoring the performance characterization of the PCM by displaying raw data and plotting spectral properties.

\subsubsection{Instruments Tab}
It is crucial to ensure all instruments are properly connected before beginning the experiment. For that reason, the Instruments tab, shown in Fig. \ref{fig:gui_overview}(c), includes information to connect and operate the oscilloscope and function generator. Options such as the resource (the instrument identifier), driver, channel, voltage range, and acquisition time are selected by the user. The Instrument tab also contains buttons for connecting to the relay and laser power supply via serial connections.

\subsubsection{Measurement Mode and Stop Condition Tabs}
The Measurement Mode tab (Fig. \ref{fig:gui_overview}(d)) and Stop Condition tab (Fig. \ref{fig:gui_overview}(f)) both allow the user to select options for terminating the experiment when a particular condition, or combination of conditions, is met. These tabs are especially useful for hours-long reliability tests. In the Stop Condition tab, the user can choose to stop the experiment when either the transmittance, CWL, or both properties fall outside a predetermined limit. Similarly, the experiment may stop after a particular number of occurrences outside that limit, or when the number of occurrences reaches a set percentage of tuning cycles completed thus far. In contrast, the conditions in the Measurement Mode tab---total cycles and total time---are calculated in the GUI itself without relying on the connected hardware. 

\subsubsection{Pulse Settings Tab}
The required energy and pulse duration may change depending on the particular PCM that is under test. In the Pulse Settings tab (Fig. \ref{fig:gui_overview}(a)), the user can set the desired laser voltages (V), pulse widths (ns), and delays (\textmu{}s) between pulses. The GUI then calculates the duty cycles (\%), frequencies (MHz), and periods (\textmu{}s) of the pulses for amorphization and crystallization. The specific values of the laser voltage, pulse width, and delay are determined experimentally and vary depending on the particular materials utilized in the tunable filter.

\subsubsection{Tabs for Monitoring Raw Data}
The top-right quadrant (\ref{fig:gui_overview}(b)) of the GUI window is for displaying ``raw'' data: in general, data that come directly from instruments, not values that are computed as a result of the instruments' data. In this quadrant, there are two tabs: the ``Laser (Osc/FGen)'' tab and the IR Image tab. As the names imply, the former plots the input to and output from the laser, as determined by the function generator and oscilloscope. The input signal to the laser differs from the output signal; the laser requires two pulses of 5-V amplitude and 10-\textmu{}s duration, with 170\,\textmu{}s between pulses. %
However, the output pulse from the laser head is only nanoseconds-long. The IR Image tab displays the IR image and ROI that are selected before the experiment. As the CWL of the PCM-based filter is tuned, the IR image will change based on the indicator on the LVF.

\subsubsection{Tabs for Testing and Performance Characterization}
The Testing tab (Fig. \ref{fig:gui_overview}(g)) contains buttons for the user to test the functionality of the GUI before running the experiment. The relays can be turned on and off with the click of a button. Sample files can also be imported, after which the IR image can be plotted, and the spectral properties can be calculated. The Files tab (Fig. \ref{fig:gui_overview}(f)) allows the user to specify file and folder names from which data will be imported and to which the data will be saved.

The bottom-right quadrant (Fig. \ref{fig:gui_overview}(e)) of the GUI window contains plots of values computed from the IR image above. The cycle number is plotted on each abscissa whereas the CWL, FWHM, and transmittance intensity are each plotted in separate tabs on the ordinate axes. 

The \textsc{Matlab}\textsuperscript{\tiny\textregistered} GUI has much utility in the current experimental system, but in future work, the GUI will be applied to (1) an experiment to investigate the PCM-based filter's reliability and (2) a compact, and custom-designed circuit to optically switch the PCM-based filter and characterize its phase changes in the NIR waveband.
\section{Applications}
Despite the current capabilities of the automated experimental system, future developments are necessary to make PCM-based filters ready for commercial and government applications. By building upon the \textsc{Matlab}\textsuperscript{\tiny\textregistered} GUI and IR camera-based characterization in this work, two additional developments are proposed. First, experiments will be conducted to evaluate the reliability and longevity of the PCM-based filter. Second, a system to characterize and tune the PCM-based filter in the NIR waveband will be developed with commonly-available and cost-effective components. In addition, the proposed system will have reduced size and weight due to a custom-designed circuit for optically switching the phase of the PCM.

\subsection{Reliability Experiment}
\begin{figure*}
    \centering
    \includegraphics[width = 0.99 \linewidth]{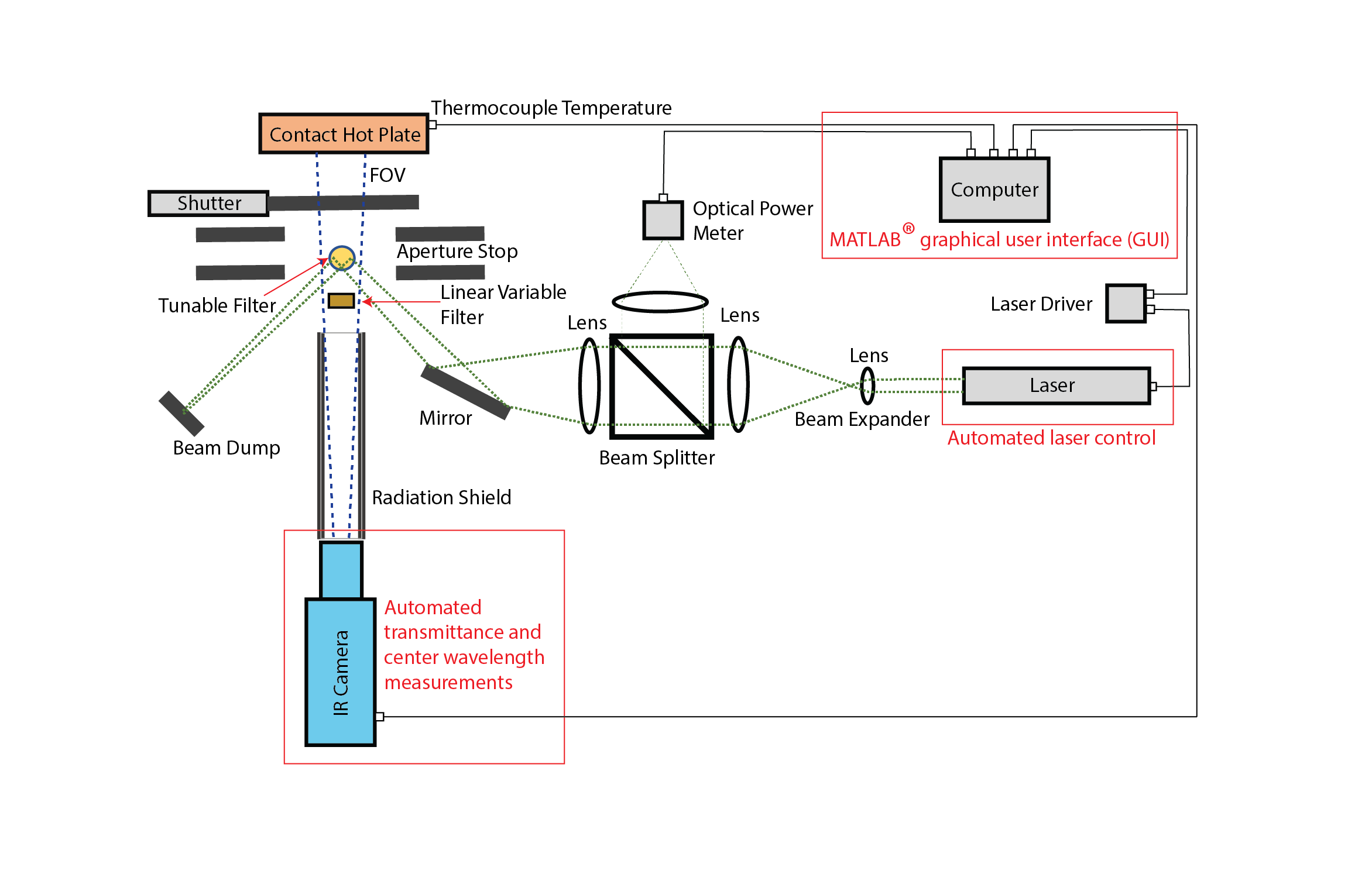}
    \caption{The reliability test setup: The reliability test will rely on the \textsc{Matlab}\textsuperscript{\tiny\textregistered} GUI for automated laser control and automated measurement of the filter's CWL. The CWL will indicate the current state of the filter.}
    \label{fig:reliability}
\end{figure*}

The reliability of the PCM-based filters will be evaluated using the \textsc{Matlab}\textsuperscript{\tiny\textregistered} GUI in future experiments. Figure \ref{fig:reliability} shows the schematic of the reliability test that will utilize the \textsc{Matlab}\textsuperscript{\tiny\textregistered} GUI. The setup is similar to that in Fig. \ref{fig:exp_setup}, but with added components to facilitate the laser-based tuning of the PCM-based filter. Because the laser will heat the PCM-based filter during the phase change, the IR camera must wait for that heat to dissipate before taking an image. This is because the IR camera cannot differentiate between surface heating of the filter and blackbody radiation. During the reliability experiment, the laser firing and the IR image acquisition will be separated for a duration that allows the heat of the PCM to sufficiently dissipate. However, the experiment will still be automated, since the filter may be potentially be switched for $10^3$--$10^6$ cycles until it fails; the moment of failure can be identified with our approach. Previous PCM reliability studies have demonstrated long operational longevity, with one example demonstrating $1.4{\times}10^{8}$ switching cycles before failure \cite{moon_reconfigurable_2019}. However, PCM-based optical filters face challenges over their resistive-based PCM counterparts. In optical PCM-based devices, the entire area of the tunable film must undergo a complete and reversible phase transition \cite{Jafari2019}. However, resistive-based PCM-based devices do not require complete crystallization of the tunable film to switch from ``high'' to ``low'' \cite{Jafari2019}.

To be used in NASA's Space and Science missions, the filter must be reliable, with any drop-off in performance as a function of the tuning cycle known in advance. For example, one potential application of the tunable filter would be the Space Launch System (SLS) for the Artemis-1 Mission. The filter could enable accurate temperature measurement of the rocket's core booster stage via radiometric techniques because of its nanosecond-scale filter switching speeds and narrow bandwidth. If the error margin of the temperature measurement is reduced, the maximum temperatures from the rocket may be smaller than originally designed for, therefore, the heat shield may not need to be as massive. The tunable filter could also be utilized for chemical and gas sensing in NASA's Science missions. If the mission time is one year, for example, the tunable filter must work reliably for that duration.

\subsection{Cost-Effective and Compact Tuning System}
\begin{figure*}
    \centering
    \includegraphics[width = 0.99 \linewidth]{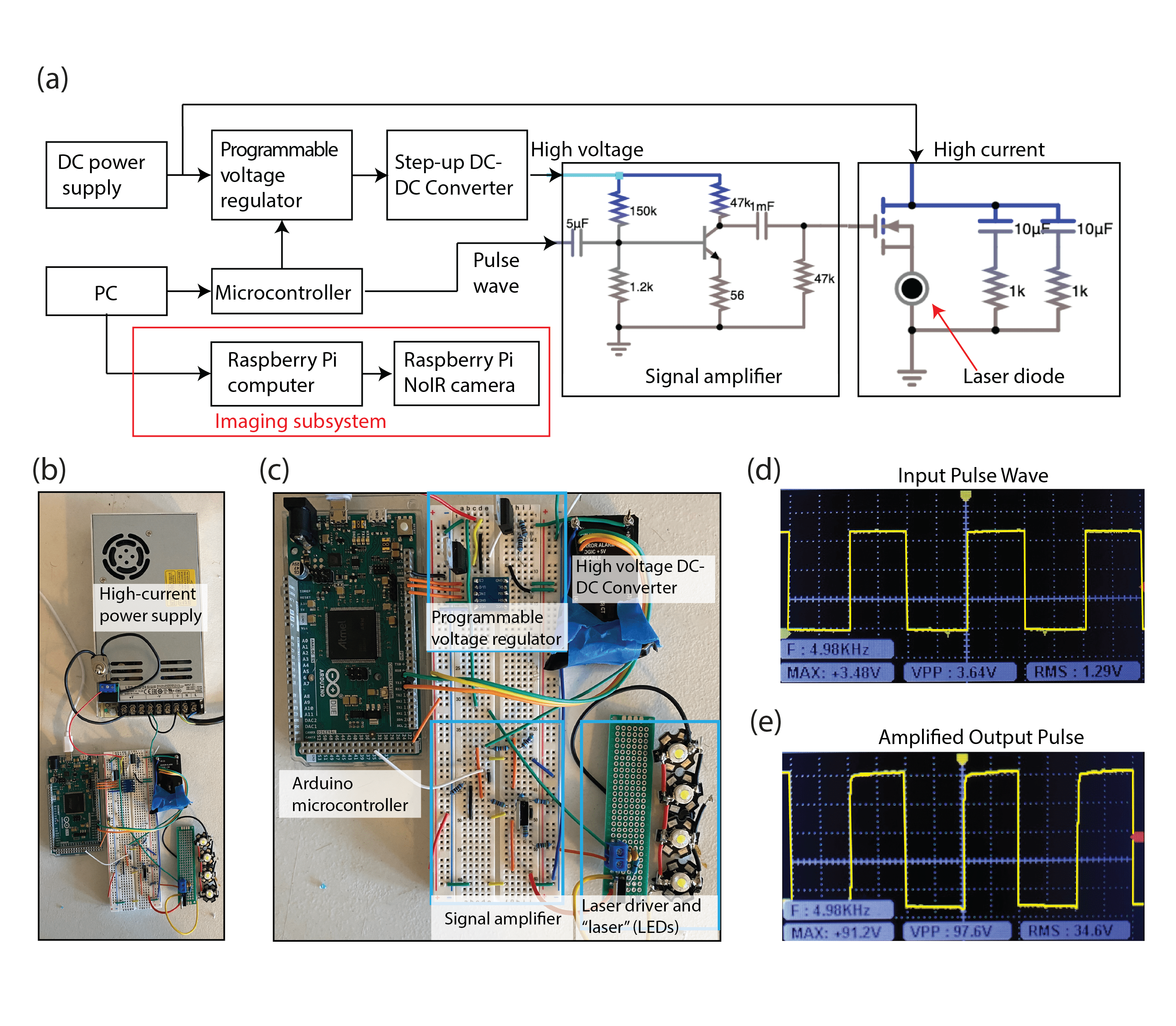}
    \caption{The cost-effective and compact tuning system. (a) The schematic of the proposed system. (b) A prototype and (c) the wired connections on the breadboard. (d) The system takes an input pulse wave from the Arduino microcontroller. (e) The output pulse waveform is amplified nearly 30 times compared to the input pulse, with minimal distortion.}
    \label{fig:low-cost}
\end{figure*}

Despite the advantages of PCM-based filters, there is a need for decreased size, weight, power, and cost; this would enlarge the range of possible applications for the tunable filter. Figure \ref{fig:low-cost} shows the design of a cost-effective and compact tuning system for PCM-based filters that have detectable phase transitions in the NIR waveband, albeit with typically an increased absorption. The cost of components for this system was less than \$400. The schematic is shown in Fig. \ref{fig:low-cost}(a), the photographs of the initial build are shown in Fig. \ref{fig:low-cost}(b)--(c), and the comparison of input and output signals are shown in Fig. \ref{fig:low-cost}(d)--(e). The system consists of two main subsystems: the phase-switching subsystem and the imaging subsystem. 

\subsubsection{Phase-Switching Subsystem}
In Fig. \ref{fig:low-cost}(a), the DC power supply (LRS-350-24, MEAN WEll Enterprises Co., Ltd.) provides power to the overall system and allows the laser diode to emit high-current pulses. A programmable voltage regulator determines the input to the step-up DC-DC converter (FS02-12, XP Power). The output of the voltage regulator (LM317, Texas Instruments) is adjusted using a digital potentiometer (X9C104). The amplitude of the DC-DC converter output determines the amplitude of the laser pulse.

The microcontroller (Due, Arduino) functions as both a waveform generator and controller for the digital potentiometer. A common-emitter amplifier circuit is utilized to amplify the pulse wave from the Arduino. In Fig. \ref{fig:low-cost}, the values of the resistors and capacitors are chosen based on circuit simulations. It is noted that other amplification strategies may be used, such as an operational amplifier with a high slew rate. A previous project created an arbitrary waveform generator using an Arduino Due and obtained pulse widths as low as 12\,ns \cite{evans_arduino_2017}. Similarly, the Arduino in Fig. \ref{fig:low-cost} was programmed to output a wide range of pulse widths and frequencies. The output from the signal amplifier determines the pulse width and frequency of the laser diode's output. Although the prototype was constructed on a breadboard, the system may be constructed on a printed circuit board (PCB) to for more compactness. This prototype was tested using one-Watt LEDs to approximate the behavior of a high-powered laser diode.The components for the prototype are available from popular retailers, making them accessible for both research and education purposes.

Preliminary results from the prototype are shown in Fig. \ref{fig:low-cost}(d)--(e). Figure \ref{fig:low-cost}(d) shows the input pulse wave from the microcontroller whereas Fig. \ref{fig:low-cost}(e) shows the output of the DC-DC converter. The input signal has a frequency of 4.98\,kHz and a peak-to-peak voltage of 3.64\,V. The output signal retained the same frequency as the input signal, but its peak-to-peak voltage was 97.6\,V. As seen in Fig. \ref{fig:low-cost}(d)--(e), the signal is minimally distorted after amplification.

\subsubsection{Imaging Subsystem}
For the imaging subsystem, a single-board computer (Model 3 B+, Raspberry Pi) is proposed to interface with an NIR camera (NoIR Camera, Raspberry Pi). This camera is similar to the Raspberry Pi V2 Camera Module, except the NoIR camera possesses no IR-cutoff filter. The spectral response of the Raspberry Pi V2 Camera Module has previously been studied \cite{pagnutti_laying_2017},and its CMOS image sensor (Si-detector) has a spectral response from approximately 400--1000\,nm. The utility in the Raspberry Pi NoIR camera is in imaging PCMs that exhibit index modulation in the NIR as well as the short-wave infrared (SWIR) and MWIR. Thus, phase transitions can be inferred based on the NIR intensity changes. Using the NIR camera, Fresnel reflections of the PCM at normal incidence would indicate the phase transitions, due to the PCM's changing refractive index upon switching. The Raspberry Pi computer and camera are directly controllable using the \textsc{Matlab}\textsuperscript{\tiny\textregistered} Raspberry Pi Support Package, making their integration into the current \textsc{Matlab}\textsuperscript{\tiny\textregistered} GUI straightforward.
\section{Discussion and Conclusion}
PCM tunable filters are emerging as an alternative technology platform for ultrafast spectral filtering across a wide range of operating wavelengths. However, the ability to accurately and easily characterize their performance remains challenging. In this study, a system for automated tunable filter characterization was designed and implemented as an alternative to FTIR spectroscopy. The LVF was shown to accurately indicate the CWL of the bandpass filter behind it. A \textsc{Matlab}\textsuperscript{\tiny\textregistered} GUI was designed to automatically control the subsystem instrumentation, display the filters' spectral properties, and record the output data. The overall system enables 1D spatiospectral (bright field) characterization and allows for easier integration of other tuning mechanisms in comparison to FTIR systems. Plans were detailed for a future reliability experimental system that is compact, low-cost, and utilizes a NIR camera. %

Future work will build upon the \textsc{Matlab}\textsuperscript{\tiny\textregistered} application and experimental results to advance the state of the art in PCM-based optical filters. Reliability studies in the future will reveal the filter's long-term performance. The compact system for imaging and switching of the filter will decrease the cost of optical PCM-based devices.
This work has wider applications in systems that utilize spatially-varying optical filters, such as circular variable filters. The software that was developed in this work can also be modified to incorporate more-sophisticated image processing algorithms, such as an algorithm to automatically detect the location of the LVF and determine the ROI. For devices and materials with more complicated spatially-varying spectral profiles, machine learning techniques may be used for more robust spectral characterization. The \textsc{Matlab}\textsuperscript{\tiny\textregistered} application is available at \url{https://dbombara.github.io/automated-pcm-characterization}.

\subsection{Paper author contributions}
D.B. developed the \textsc{Matlab}\textsuperscript{\tiny\textregistered} application. H.J.K and C.W. conceived the idea. H.J.K. and S.B. carried out the experiments. C.W. advised the device characterization. All authors wrote the manuscript with D.B. as the lead.

\subsection{Acknowledgments}
D.B. would like to thank the NASA Internship, Fellowship, and Scholarship (NIFS) program coordinators and/or project manager for the internship opportunity. The authors would also like to thank Scott Bartram, Equipment Specialist at the NASA Langley Research Center, for his expertise and assistance with laboratory instruments. Beginning in June 2020, this work has been completed under the NIFS program with the project entitled ``LabVIEW controlled evaluation system of actively tunable filter for NASA Science \& Space Mission''.

\subsection{Funding statement}
Langley Research Center (CIF, IRAD); Engineering and Physical Sciences Research Council (EP/R003599/1); Wellcome Trust (Interdisciplinary Fellowship).

\subsection{Competing interests}
The authors declare that they have no competing interests.


\vspace{2ex} \noindent \textbf{David Bombara} is a master's student in mechanical engineering at the University of Nevada, Reno. He is also an intern since June 2020 for the NASA Langley Research Center through the NIFS (NASA Internship, Fellowship, and Scholarship) program. He obtained his bachelor's degree in mechanical engineering from the University of Nevada, Reno in 2020. His broad research interests include optical instrumentation, measurement, and control systems.

\vspace{2ex} \noindent \textbf{Dr. Calum Williams} is a Research Fellow in the Department of Physics at the University of Cambridge. His research focuses on miniaturized optical imaging technologies for biomedical diagnostics. He completed his Ph.D. in nanophotonics and holography at the University of Cambridge in 2017.

\vspace{2ex} \noindent \textbf{Stephen E. Borg} is an aerospace technologist at the NASA Langley Research Center. He received bachelor's degrees in mechanical engineering (1988) and physics (1988) from Old Dominion University and has been working primarily in the areas of radiometry, infrared imaging, and temperature measurement in support of NASA's subsonic \& hypersonic aeronautics research programs.

\vspace{2ex} \noindent \textbf{Dr. Hyun Jung Kim} is an Associate Research Fellow at the National Institute of Aerospace, working at the NASA Langley Research Center (LaRC) from 2009. She received her Ph.D. in Materials Science and Engineering at the KAIST. She works on solid-state tunable filter and telescope lenses, and is the PI of LaRC's ISO5 Optical Cleanroom. Her research interests include film process and new material development to enable the next generation of optics, sensors, and energy devices.
\end{spacing}
\end{document}